\documentclass[reprint,aps,pra]{revtex4-1}
\usepackage{graphicx}
\usepackage{dcolumn}
\usepackage{bm}
\usepackage{amsmath}
\usepackage{amsthm}
\usepackage{amssymb}
\usepackage{float}
\usepackage{hyperref}
\usepackage{color}
\usepackage{epstopdf}
\usepackage{cleveref}
\usepackage[svgnames]{xcolor}
\hypersetup{hidelinks,colorlinks=true,allcolors=DarkBlue}

\newcommand{\iNt}{\mathrm{int}}

\begin{document}

\preprint{APS/123-QED}

\title{Mutual information--energy inequality \\ for thermal states of a bipartite quantum system}
\author{Aleksey Fedorov$^{1,2}$}\email{akf@rqc.ru}
\author{Evgeny Kiktenko$^{2,3}$}\email{evgeniy.kiktenko@gmail.com}
\affiliation
{
\mbox{$^{1}$Russian Quantum Center, Skolkovo, Moscow 143025, Russia}
\mbox{$^{2}$Bauman Moscow State Technical University, Moscow 105005, Russia}
\mbox{$^{3}$Geoelectromagnetic Research Center of Schmidt Institute of Physics of the Earth,}
\mbox{Russian Academy of Sciences, Troitsk, Moscow Region 142190, Russia}
}

\date{\today}

\begin{abstract}
In this work, we consider an upper bound for the quantum mutual information in thermal states of a bipartite quantum system.
This bound is related with the interaction energy and logarithm of the partition function of the system.
We demonstrate the connection between this upper bound and the value of the mutual information for the bipartite system realized by two spin-1/2 particles in the external magnetic field with the ${\rm XY}$-Heisenberg interaction.
\begin{description}
\item[PACS numbers]
03.65.Wj, 03.65.-w, 03.67.-a
\end{description}
\end{abstract}
                              
\maketitle

\section{Introduction}
A paradigm of consideration of quantum systems as a potential basis for applications in various technologies, {\it e.g.}, information processing, communications, and metrology, 
inspires a new wave of studies many-body systems with focus on quantum correlation, quantum entanglement, and discord phenomena \cite{Vedral,Vedral2,Vedral3}.
In particular, outstanding findings are obtained for spin models 
\cite{Lieb,Divincenzo,Kane,Divincenzo2,Divincenzo3,Sorensen,Cirac,Lidar,Divincenzo4,Imamoglu,Privman,Zheng,Wang,Franchini,Franchini2,Lychkovskiy,Huang,Huang2}. 
Remarkable progress in experiments with electronic spins \cite{Divincenzo}, 
nuclear spins \cite{Kane}, 
quantum dots \cite{Divincenzo2,Divincenzo3}, 
and atomic ensembles in optical lattices \cite{Sorensen,Cirac}, allows to design quantum systems with effective spin chain Hamiltonians. 
This fact opens intriguing prospectives for investigation of fundamental concepts in these systems and revealing novel interesting phenomena.

Paramount importance of the ${\rm XY}$ Hamiltonian spin model \cite{Lieb} has been demonstrated \cite{Imamoglu,Privman,Zheng,Wang,Franchini,Franchini2}. 
In particular, it has been shown that ${\rm XY}$ Hamiltonian appears in quantum dot spins and cavity QED systems towards to their applications for quantum information processing 
\cite{Imamoglu,Privman,Zheng,Lidar,Divincenzo4}. 

Recently, a novel class of so-called entropy--energy inequalities of has been revealed \cite{Manko0}.
This class of inequalities connects the mean value of the Hamiltonian and the entropy of an arbitrary state for a given Hamiltonian of the system. 
In the present work, we follow the strategy of Ref.~\cite{Manko0} to construct an upper bound for the quantum mutual information in thermal states of a bipartite quantum system.
We consider the relation between this bound and the value of the mutual information for the bipartite system realized by two spin-1/2 particles in the external magnetic field with the ${\rm XY}$-Heisenberg interaction.

Our paper is organized as follows. 

In Section \ref{inequality}, we construct the inequality for the quantum mutual information and energy in bipartite quantum systems.
We consider this inequality in the framework of the model of spin-1/2 particles with the XY Hamiltonian in the external magnetic field in Section \ref{XY}.
In Section \ref{conclusion}, we give our conclusion and prospectives. 

\section{Mutual information--energy inequality}\label{inequality}

\begin{figure*}[t]
\begin{centering}
\includegraphics[width=0.8\linewidth]{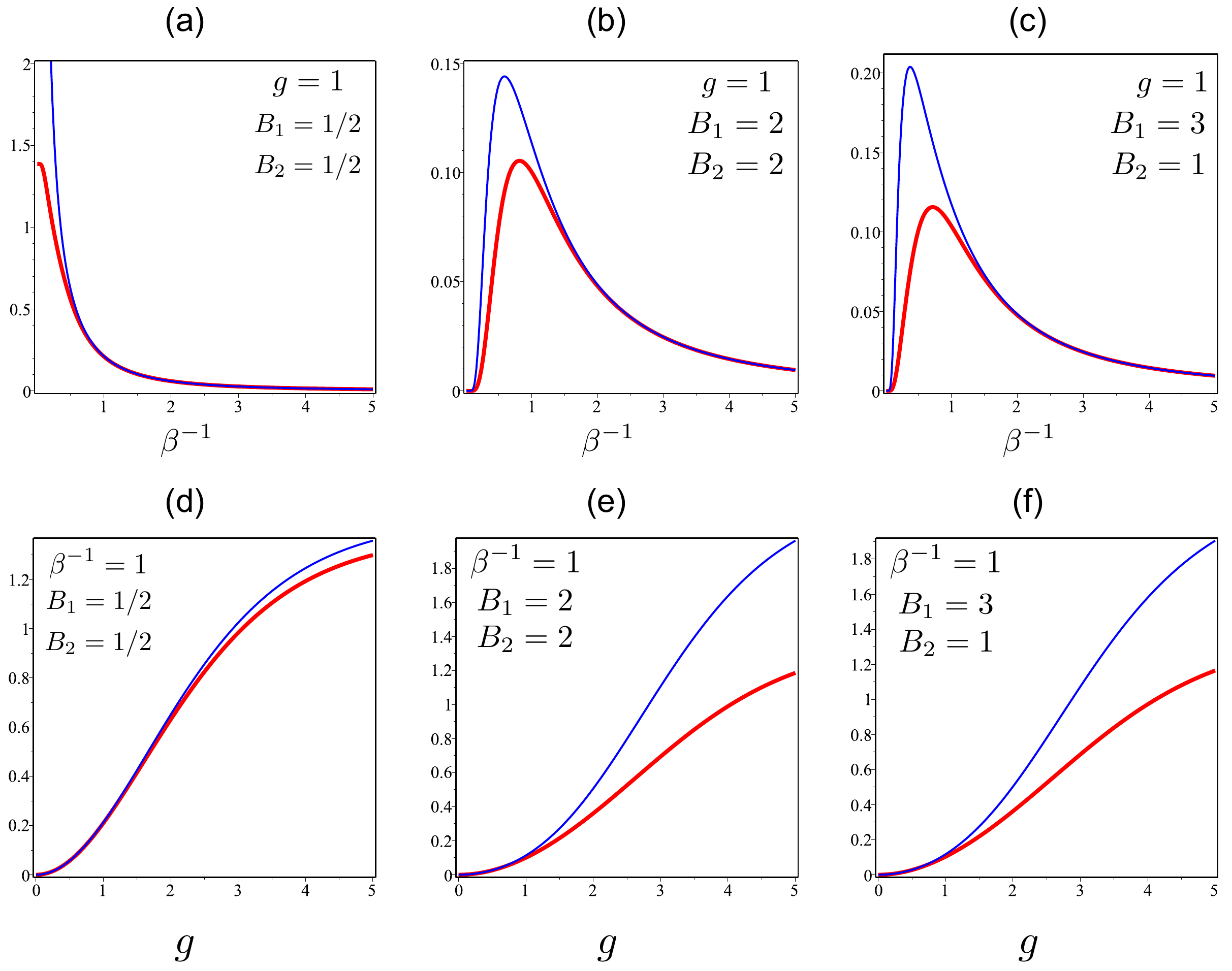}
\end{centering}
\caption
{
Quantum mutual information (red) given by  Eq. (\ref{eq:MI}) and its upper bound (blue) given by Eq. (\ref{eq:UB}):
(a)--(c) shown as functions of temperature $\beta^{-1}$ at fixed coupling constant $g=1$; 
(d)--(f) shown as functions of the coupling constant $g$ at fixed temperature $\beta^{-1}=1$. 
The external magnetic fields are: 
(a) and (d) $B_1=B_2=1/2$; 
(b) and (e) $B_1=B_2=2$; 
(c) and (f) $B_1=3, B_2=1$.
}
\label{fig:1}
\end{figure*}

Consider a bipartite quantum system $AB$, which is described by density operator in the Hilbert space $\mathcal{H}_{AB}=\mathcal{H}_{A}\otimes \mathcal{H}_{B}$.
The Hamiltonian of the system has the following generic form:
\begin{equation}\label{eq:Ham}
	H=H_A\otimes{\mathbb I}_B +{\mathbb I}_A\otimes H_B+H_\iNt,
\end{equation}
where $H_A$ and $H_B$ describe the subsystems without an interaction, 
$H_\iNt$ describes the coupling between subsystems, 
${\mathbb I}_A$ and ${\mathbb I}_B$ are the identity operators in $\mathcal{H}_{A}$ and $\mathcal{H}_{B}$ respectively.

We study the state of thermal equilibrium, which is given by the following expression:
\begin{equation}\label{eq:ThSt}
	\rho_{AB}=\frac{1}{Z_{AB}}\exp{(-\beta H)}, 
	\quad 
	Z_{AB}={\rm Tr}[\exp{(-\beta{H})}],
\end{equation}
where $\beta$ is inverse temperature in the corresponding energy units and $Z$ is the partition function.
Taking partial traces in Eq. (\ref{eq:ThSt}), we obtain the states of subsystem:
\begin{equation}
	\rho_A={\rm Tr}_B[\rho_{AB}], 
	\quad 	
	\rho_B={\rm Tr}_A[\rho_{AB}].
\end{equation}
It is clear that the total energy of state (\ref{eq:ThSt}) could be represented as a sum of three terms:
\begin{equation}
	\begin{split}
		&E=E_A+E_B+E_\iNt,
		\quad 
		E_A={\rm Tr}[\rho_A H_A], 
		\\ 	
		&E_B={\rm Tr}[\rho_B H_B ],
		\quad 
		E_\iNt={\rm Tr}[\rho_{AB} H_\iNt].
	\end{split}	
\end{equation}

Following the strategy of Ref. \cite{Manko0}, to derive an inequality between the mutual information and energy in bipartite thermal state, we start with the notation of the relative entropy \cite{Vedral},
\begin{equation}\label{eq:RelEntr}
	S(\rho||\sigma)={\rm Tr}[\rho\ln\rho-\rho\ln\sigma]\geq0.
\end{equation}
This relation is a measure of difference between states with density matrices $\rho$ and $\sigma$.
We note that the identity $S(\rho\|\sigma)=0$ takes place in the case $\rho=\sigma$ only.
Using the von Neumann entropy
\begin{equation}
	S(\rho)=-{\rm Tr}[\rho\ln\rho],
\end{equation}
one can rewrite Eq. (\ref{eq:RelEntr}) as follows:
\begin{equation} \label{eq:RelEntr2}
	S(\rho\|\sigma)=-S(\rho)-{\rm Tr}[\rho\ln\sigma].
\end{equation}

By expanding the trivial identity $S(\rho_{AB}||\rho_{AB})=0$ and substituting the explicit form of thermal state (\ref{eq:ThSt}) in second term in the RHS of Eq. (\ref{eq:RelEntr2}), we obtain:
\begin{equation}
	\begin{split}	
		0=&-S(\rho_{AB})-{\rm Tr}[\rho_{AB}(-\beta H-\ln Z_{AB})]= \\
		&-S(\rho_{AB})+\beta E+\ln Z_{AB},
	\end{split}	
\end{equation}
or, equivalently, 
\begin{equation}\label{eq:SAB}
	S(\rho_{AB})=\beta E+\ln Z_{AB}.
\end{equation}

Let us consider the relative entropy between the state $\rho_A$ of the subsystem $A$ and an another thermal state $\tilde{\rho}_A$,
\begin{equation}
	\tilde{\rho}_A=\frac{1}{Z_A}\exp{(-\beta H_A)}, 
	\quad 
	Z_{A}={\rm Tr}[\exp{(-\beta H_A)}].
\end{equation}
In the general case, we have
\begin{equation}
	\begin{split}
		0&\leq S(\rho_A||\tilde{\rho}_A)= \\
		&-S(\rho_A)-{\rm Tr}[\rho_{A}(-\beta H_A-\ln Z_{A})= \\
		&-S(\rho_{A})+\beta E_A+\ln Z_{A}
	\end{split}	
\end{equation}
or, equivalently, the following inequality:
\begin{equation} \label{eq:SA}
	S(\rho_{A})\leq \beta E_A+\ln Z_{A}.
\end{equation}
One can see that the same inequality could be obtained for the entropy of $\rho_B$:
\begin{equation}\label{eq:SB}
	S(\rho_{B})\leq \beta E_B+\ln Z_{B},
	\quad
	Z_{B}={\rm Tr}[\exp{(-\beta H_B)}].
\end{equation}

Finally, by considering the standard definition of the quantum mutual information
\begin{equation} \label{eq:MI}
	I(\rho_{AB})=S(\rho_A)+S(\rho_B)-S(\rho_{AB})
\end{equation}
and using Eqs. (\ref{eq:SAB}), (\ref{eq:SA}), and (\ref{eq:SB}), we obtain the following inequality:
\begin{equation} \label{eq:Inf}
	\begin{split}
	I(\rho_{AB})&\leq\beta(E_A+E_B-E_{AB})+\ln\frac{Z_AZ_B}{Z_{AB}}= \\
	&-\beta E_\iNt+\ln\frac{Z_AZ_B}{Z_{AB}},
	\end{split}	
\end{equation}
which gives the upper bound for mutual information in thermal bipartite state.

For the purposes of convenience, we introduce the notation for the upper bound of the mutual information
\begin{equation} \label{eq:UB}
	I_{\rm ub}(\rho_{AB})=-\beta E_\iNt+\ln\frac{Z_AZ_B}{Z_{AB}}.
\end{equation}
Clearly that in the case of absence of interaction, {\it i.e.}, at $H_\iNt=0$, the upper bound and mutual information itself goes to zero:
\begin{equation}
	Z_{AB}=Z_AZ_B, 
	\quad 
	I(\rho_{AB})=I_{\rm ub}(\rho_{AB})=0.
\end{equation}
Inequality (\ref{eq:Inf}) is a central object of the present paper.

\section{Results for ${\rm XY}$-Heisenberg interaction}\label{XY}

In this section, we discuss an implementation of inequality (\ref{eq:Inf}) for the particular spin model.
We consider ${\rm XY}$-Heisenberg interaction between two spin-1/2 particles in an inhomogeneous magnetic field acting along $Z$-axis.
The systems is described by the following Hamiltonian: 
\begin{equation}\label{eq:XY}
	H_A=B_1\sigma_z, 
	\quad 
	H_B=B_2\sigma_z,
	\quad 
	H_\iNt=g(\sigma_x\otimes\sigma_x+\sigma_y\otimes\sigma_y),
\end{equation}
where $\sigma_i (i\in(x,y,z)$ stands for the corresponding Pauli matrices, 
$g$ is the coupling constant, 
and $B_1$ and $B_2$ are magnetic fields acting on particles (magnetic moments of particles are set to be unit).

The ground state of Hamiltonian~(\ref{eq:XY}) turns to be the entangled state
\begin{equation}
	|\Psi\rangle=\frac{1}{\mathcal{C}}\left(\frac{2g}{\sqrt{(B_1-B_2)^2+2g^2}}|\uparrow\downarrow\rangle-|\downarrow\uparrow\rangle\right),
\end{equation}
at $B_1B_2<g^2$  with $\mathcal{C}$ being the normalization factor, and the separable state at $B_1B_2>g^2$ (for positive $B_1$ and $B_2$, the state is $|\uparrow\uparrow\rangle$, and it is $|\downarrow\downarrow\rangle$ otherwise).

We start from consideration the case with no external field, {\it i.e.}, $B_1=B_2=0$.
Due to the symmetry of Hamiltonian (\ref{eq:XY}), the reduced states $\rho_A$ and $\rho_B$ of corresponding thermal state $\rho_{AB}$ (\ref{eq:ThSt}) are proportional to identity operators.
On the other hand, we have $Z_A=Z_B=2$ (as the dimension of the Hilbert space of the subsystems) and $E_A=E_B=0$.
Thus, inequalities (\ref{eq:SA}) and (\ref{eq:SB}) turn into identities and we obtain:
\begin{equation}
	I(\rho_{AB})=I_{\rm ub}(\rho_{AB})=-\beta E-\ln{Z_{AB}}+2\ln{2}.
\end{equation}

Situations of non-zero uniform and nonuniform magnetic fields are presented in Fig.~\ref{fig:1}.
First we note that in Fig.~\ref{fig:1}a, with entangled ground state, the upper bound~\eqref{eq:UB} explodes at $\beta^{-1}\rightarrow 0$ and becomes even higher than maximal level of mutual information for two-qubit system, {\it i.e.}, the value $2\ln 2$.
However, at high temperatures $\beta^{-1}>g$ the upper bound tends to real value of the mutual information at all considered values of the magnetic field.

The behavior of the quantum mutual information and its upper bound as functions of the coupling constant $g$ at fixed temperature $\beta^{-1}$, which is presented in Fig.~\ref{fig:1}(d-f), 
confirms with foregoing statement: the difference between $I(\rho_{AB})$ and $I_{\rm ub}(\rho_{AB})$ grows with increasing of the coupling constant $g$.
It seems like non-uniformity of magnetic field does not bring any significant changes in the general picture.

\section{Conclusion}\label{conclusion}

We have introduced the upper bound (\ref{eq:UB}) for the quantum mutual information in thermal states of bipartite quantum systems based on the interaction energy and logarithm of the partition functions.
To illustrate our results, we have considered two spin-1/2 particles interacting by ${\rm XY}$-Heisenberg model in the inhomogeneous magnetic field.
It turned out that in the considered model the introduced the upper bound tends to the value of mutual information for the values of temperatures (in the energy units) higher than the interaction energy.
The important open question is whether it is true for all possible models of interactions and dimensions of subsystems.

{\bf Acknowledgments}.
Authors thank V.I. Man'ko and O.V. Man'ko for productive discussions.
We thank S.N. Filippov, Y.V. Kurochkin, and Y. Huang for useful comments. 
We acknowledge the support from the 
Council for Grants of the President of the Russian Federation (grant SP-961.2013.5, EOK), 
the Dynasty Foundation (AKF), and RFBR (projects 14-08-00606 \& 14-02-00937).
We are also grateful to the organizers of the 23rd Laser Physics Workshop (Sofia, July 14--18, 2014) for kind hospitality.

\end{document}